\documentclass[prx,twocolumn,showpacs,preprintnumbers,amsmath,amssymb,superscriptaddress]{revtex4}
\usepackage[usenames]{color}

\usepackage{graphicx}
\usepackage{bm}

\begin{document}

\title{Linear response theory for electron-hole pair kinetics: exciton formation}
\author{Shota Ono}
\email{shota\_o@gifu-u.ac.jp}
\affiliation{Department of Electrical, Electronic and Computer Engineering, Gifu University, Gifu 501-1193, Japan}
\affiliation{Department of Physics, Graduate School of Engineering,Yokohama National University, Yokohama, 240-8501, Japan}
\affiliation{Department of Physics, University of California, Berkeley, California 94720-7300, USA}
\date{\today}
\begin{abstract}
A linear response theory for electron-hole pair density is developed, which constitutes a new theoretical method, and a definition of exciton density in a first-principles context is derived by considering both the electron-hole attractive interaction and the screening effect. This allows the exciton time evolution to be examined. This formulation is applied to a jellium model in order to prove the existence of a transient exciton, and to observe crossover from a transient to a stable exciton in response to decreased electron density. The exciton formation mechanism is also revealed.
\end{abstract}

\maketitle
\section{Introduction}
The exciton is one of the most important elementary excitations in condensed matter. Intense peaks below the single-particle absorption edge of the optical absorption spectra of molecules, semiconductors, and insulators indicate the presence of these particles \cite{RL,onida}. This is particularly true for systems with high dielectric constants, in which the effective-mass approximation may be valid \cite{elliott}.

It is believed that exciton observation in metals is quite difficult, because free carriers immediately screen the holes created by photon absorption. Recently, however, Cui et al. have reported that excitons exist in metals in the form of {\it transient} excitons, based on findings obtained through the application of multi-photon photoemission spectroscopy to a silver surface \cite{Cui}. The transient excitonic state that does not correspond to anything in the single-particle band structure develops into an image potential state within 100 fs, which is quite short compared to the exciton lifetime in typical insulators. The findings of this experiment pose the question of how to define the existence of an exciton within a short time scale, and move the field in the direction of developing an understanding of excitons in both space and time.

The screening effect is significant as regards exciton existence. After the creation of the photohole, the Coulomb potential between the electrons and holes is gradually weakened over time \cite{schone1}. Since screening is incomplete in an insulator, the Coulomb potential approaches an approximate ratio of the bare Coulomb potential and the dielectric constant, which causes long-lived excitons to appear. On the other hand, the screening in a metal is complete. The Coulomb potential approaches the well-known Tomas-Fermi potential, which generates no bound states in general \cite{rogers}. However, Sch\"{o}ne and Ekardt have theoretically suggested the possibility of transient excitons occurring in bulk metals until screening completion, although this is a very short timescale \cite{schone2,schone3}. A similar conclusion has been reached by Gumhalter et al., via a systematic calculation for metal surfaces \cite{Gumhalter}. In both studies, the transient excitons are described by an effective-mass equation under a time-dependent potential for the electron, which is calculated using a linear response theory. In this approach, two bands relevant to the exciton formation must be chosen {\it a priori}, which may introduce an arbitrariness to the definition. The Bethe-Salpeter equation (BSE) is, in principle, an exact scheme describing the excitonic properties of electronic systems beyond the effective mass approximation \cite{ShamRice}. However, the present method of solving the BSE can be applied to a system under a stationary interaction potential only. Thus, to understand the kinetics of the exciton, it is highly desirable to develop a theory that passes beyond both the effective-mass approximation and the use of a stationary interaction potential in the BSE. Note that, although Attaccalite et al. have derived an equation of motion for the non-equilibrium Green's function that may be regarded as a time-dependent BSE \cite{marini}, the application of this theory has been limited to the calculation of the optical absorption spectra in finite systems and wide-gap semiconductors. 

In this paper, a theory that allows the kinetics of the exciton to be described is developed. This approach is based on a linear response theory for the electron-hole (EH) pair density, which allows direct computation of the time-dependent EH pair density fluctuations under an external perturbation. A natural definition of the exciton is derived by considering the EH attractive interaction and the screening effect. Application to a jellium model proves the existence of the transient exciton and reveals a novel property of the time-evolution of exciton  formation.

\section{Formulation}
We consider an EH pair density operator defined as 
\begin{eqnarray}
 n_2(x,x') =  \psi^\dagger (x') \psi (x) \psi^\dagger (x) \psi (x'),
\end{eqnarray}
where $x$ and $x'$ represent the position ($\bm{x}$ and $\bm{x}'$) and spin ($\sigma$ and $\sigma'$) of the electron and hole, respectively. The expectation value of this operator with respect to the ground state is equivalent to the EH pair density in the system
\begin{eqnarray}
 \langle N,0 \vert n_2(x,x') \vert N, 0 \rangle
&=& \sum_S \left\vert \langle N,S \vert \psi^\dagger (x) \psi (x') \vert N, 0 \rangle \right \vert^2, 
\nonumber\\
\end{eqnarray}
where $\vert N, S \rangle$ is an arbitrary excited state in the $N$-electron system. A carefully selected perturbation excites the excitons, and their decay can be studied within the framework of a linear response theory. In general, the interaction Hamiltonian between a system of charged particles and the electromagnetic field is the sum of the scalar and vector potential components, such that
\begin{eqnarray}
H' (t) = H_s' (t) + H_v'(t).
\end{eqnarray}
Then, the linear response is expressed by
\begin{eqnarray}
 \delta \langle n_2(x,x';t) \rangle 
 &=& \frac{i}{\hbar} \int dt' 
 \langle N,0 \vert [ H'(t'), n_2(x,x';t) ] \vert N, 0 \rangle.
 \nonumber\\
 \label{eq:deltan2}
\end{eqnarray}
Here, $\delta \langle n_2(x,x';t) \rangle$ describes the induced EH pair density at time $t$ (an electron and hole at $x$ and $x'$, respectively) caused by an external perturbation at $t'$. As a simple example, we consider a scalar potential
\begin{eqnarray}
H' (t) = - \int dx e\varphi^{\rm ext} (\bm{x};t) n_1(x;t),
\end{eqnarray}
where $e$ is the charge, $\varphi^{\rm ext} (\bm{x};t)$ is the time-dependent and spin-independent scalar potential, and $n_1(x;t)$ is the electron density operator in the Heisenberg picture, $n_1(x;t) = e^{iHt/\hbar} \psi^\dagger (x) \psi (x) e^{-iHt/\hbar}$ ($H$ is the unperturbed Hamiltonian). 
If the retarded correlation function is defined by
\begin{eqnarray}
& & D_{2}^{\mathrm{R}}(x,x',x'';t-t') \nonumber\\
&=& -i \theta(t-t') 
 \langle N,0 \vert [ n_2(x;x';t), n_1(x'';t') ] \vert N, 0 \rangle,
\end{eqnarray}
Eq.~(\ref{eq:deltan2}) may be rewritten as
\begin{eqnarray}
 \delta \langle n_2(x;x';t) \rangle 
 &=& -\frac{e}{\hbar} \int dx'' \int_{-\infty}^{\infty} dt' \nonumber\\
 &\times& \varphi^{\rm ext} (\bm{x}'';t') D_{2}^{\mathrm{R}}(x,x',x'';t-t').
\end{eqnarray}
The computation of $\delta \langle n_2(x,x';t) \rangle$ enables us to derive the lifetimes of the excitons in the system in question under a certain excitation. Here, we consider a homogeneous system whose Hamiltonian is written as (see Ref.~\cite{fetter_walecka})
\begin{eqnarray}
 H &=& H_0 + H_1 \nonumber\\
     &=& \int dx \psi^{\dagger} (x) \left ( - \frac{\hbar^2}{2m} \nabla^2 \right) \psi (x) \nonumber\\
 &+& \frac{1}{2} \int dx \int dx' \psi^{\dagger} (x)\psi^{\dagger} (x') V(\bm{x}-\bm{x}') \psi (x') \psi (x), 
 \label{eq:totalH}
 \nonumber\\
\end{eqnarray}
where $H_0$ is the non-interacting Hamiltonian describing the kinetic energy of the electron systems and $H_1$ is the perturbation Hamiltonian describing the interaction energy between the electrons at position $\bm{x}$ and $\bm{x}'$ via $V(\bm{x}-\bm{x}')$. In this system, $D_{2}^{\rm R} (x,x',x'';t-t')$ is a function of $\bm{x}-\bm{x}''$ and $\bm{x}'-\bm{x}''$, and the Fourier transformation is defined as 
\begin{eqnarray}
 \tilde{D}_{2}^{\mathrm{R}} (q,q',\sigma'';\omega) 
 &=& \int d (\bm{x}-\bm{x}'') \int d (\bm{x}'-\bm{x}'')  \nonumber\\
&\times& e^{-i\bm{q}\cdot (\bm{x}-\bm{x}'')}  e^{-i\bm{q}'\cdot (\bm{x}'-\bm{x}'')} \nonumber\\
&\times& \int d (t-t') e^{i\omega (t-t')} \nonumber\\
&\times& D_{2}^{\mathrm{R}} (x,x',x'';t-t'),
\end{eqnarray}
where $q=(\bm{q},\sigma)$ denotes the wavevector and spin. The time-dependence of the EH pair density in the momentum space is expressed as
\begin{eqnarray}
\delta \langle \tilde{n}_2(q,q';t) \rangle 
 &=& \sum_{\sigma''}\int \frac{d\omega}{2\pi} e^{- i\omega t} 
 \frac{(-e)}{\hbar} \tilde{\varphi}^{\rm ext} (\bm{q}+\bm{q}';\omega) \nonumber\\
 &\times&\tilde{D}_{2}^{\mathrm{R}}(q,q',\sigma'';\omega).
 \label{eq:n2qq}
\end{eqnarray}
Equation (\ref{eq:n2qq}) can also be extended to treat the reciprocal lattice vector, which enables us to study periodic systems. To evaluate $\tilde{D}_{2}^{\mathrm{R}}(q,q',\sigma'';\omega)$, we first consider a time-ordered correlation function
\begin{eqnarray}
 & &D_{2}^{\mathrm{T}}(x,x',x'';t-t') \nonumber\\
 &=& -i \langle N,0 \vert {\mathrm T}[ n_2(x,x';t) n_1(x'';t') ] \vert N, 0 \rangle.
 \label{eq:TOC}
\end{eqnarray}
The perturbation expansion method can be used, because of the presence of the time-ordering operator ${\mathrm T}$ in Eq.~(\ref{eq:TOC}). The Fourier transformation of the time-ordered correlation function of $D_{2}^{\rm T}$ for $q$ and a frequency ($\omega$) space is given as, respectively,
\begin{eqnarray}
 \tilde{D}_{2}^{\mathrm{T}} (q,q',\sigma'';\omega) 
 &=& \int d (\bm{x}-\bm{x}'') \int d (\bm{x}'-\bm{x}'') \nonumber\\
 &\times& e^{-i\bm{q}\cdot (\bm{x}-\bm{x}'')}  e^{-i\bm{q}'\cdot (\bm{x}'-\bm{x}'')} \nonumber\\
 &\times& D_{2}^{\mathrm{T}} (x,x',x'';\omega), 
\end{eqnarray}
and 
\begin{eqnarray}
 D_{2}^{\mathrm{T}}(x,x',x'';\omega) &=& \int d(t-t') e^{i\omega (t-t')} \nonumber\\
 &\times& D_{2}^{\mathrm{T}} (x,x',x'';t-t').
 \nonumber\\
\end{eqnarray}
Using the following relations in $\omega$ representation: $D_{2}^{\mathrm{T}}(x,x',x'';\omega) = D_{2}^{\mathrm{R}}(x,x',x'';\omega)$ for $\omega>0$ and $D_{2}^{\mathrm{R}}(x,x',x'';-\omega) = D_{2}^{\mathrm{R}*}(x,x',x'';\omega)$, one can obtain $D_{2}^{\mathrm{R}}(x,x',x'';\omega)$ for all values of $\omega$ (Ref.~\cite{strinati}). Finally, $\tilde{D}_{2}^{\mathrm{R}}(q,q',\sigma'';\omega)$ can be obtained after Fourier transforming the space dependence. 

\begin{figure}[t]
\center
\includegraphics[scale=0.35,clip]{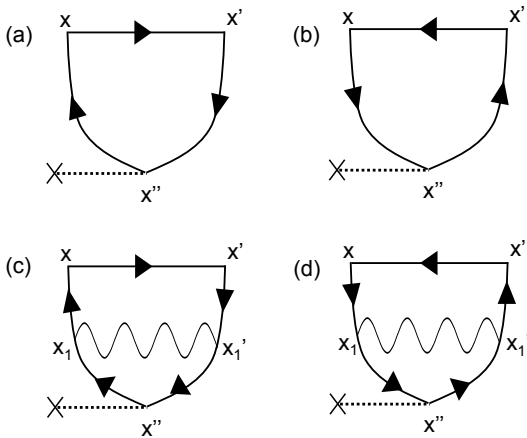}
\caption{\label{fig:diagram}
Contribution to $D_{2}^{\rm T}(x,x',x'';t-t')$. (a) and (b) are  zeroth order, while (${\mathrm c}$) and (d) show the first-order contribution. The cross and the dotted line indicate the external perturbation, and the wavy line indicates the Coulomb interaction.}
\end{figure}

\section{Exciton description}
To study the exciton existence, we must consider the competing effect between the electron-hole attractive interaction and the screening, because the former determines the two-particle trend, while the latter gives rise to the single-particle behavior. 

We study the attractive interaction using the perturbation expansion of $D_{2}^{\rm T}(x,x',x'';t-t')$ given by Eq.~(\ref{eq:TOC}). The evaluation of $D_{2}^{\mathrm{T}}(x,x',x'';\omega)$ is performed by using the non-interacting Green's function
\begin{eqnarray}
& &  G^{(0)}(x,x';\omega) \nonumber\\
&=& \delta_{\sigma\sigma'} G_{\sigma}^{(0)}(\bm{x},\bm{x}';\omega) \nonumber\\
&=& \delta_{\sigma\sigma'} 
\left(
   \sum_{\alpha}^{\rm occ} 
   \frac{\phi_{\alpha\sigma}(\bm{x}) \phi_{\alpha\sigma}^{*}(\bm{x}')}{\omega - \omega_{\alpha} - i\delta}
+ \sum_{\beta}^{\rm emp} 
    \frac{\phi_{\beta\sigma}(\bm{x}) \phi_{\beta\sigma}^{*}(\bm{x}')}{\omega - \omega_{\beta} + i\delta}
 \right),
 \label{eq:G0}
 \nonumber\\
\end{eqnarray}
where $\phi_{\alpha\sigma}(\bm{x})$ denotes the single-particle eigenfunction with quantum number $\alpha$ and spin $\sigma$ for the non-interacting Hamiltonian. The product of $\omega_\alpha$ and the Planck constant $\hbar$ is the energy of the single-particle state $\alpha$. ''occ'' and ''emp'' denote the occupied states $\alpha$ and empty states $\beta$, respectively. The fact that the non-interacting Hamiltonian does not change the spin of the electron and hole is assumed. In the homogeneous system described by Eq.~(\ref{eq:totalH}), the single-particle eigenfunction for $H_0$ is expressed by the plane waves 
\begin{eqnarray}
 \phi_{\alpha\sigma} (\bm{x}) = \frac{1}{\sqrt{\Omega}} e^{i\bm{k}_\alpha \cdot \bm{x}} s(\sigma),
 \label{eq:pw}
\end{eqnarray} 
where $\bm{k}_\alpha$ is the wavevector, $s(\sigma)$ is the spin function, and $\Omega$ is the volume of the system. Then, the summation in Eq.~(\ref{eq:G0}) is replaced by the following integral
\begin{eqnarray}
 \sum_{\alpha}^{\mathrm{occ}} &=& \int d\bm{k}_\alpha \frac{\Omega}{(2\pi)^3} 
 \theta_H (k_F - \vert \bm{k}_\alpha \vert), \\
 \sum_{\beta}^{{\rm emp}} &=& \int d\bm{k}_\beta \frac{\Omega}{(2\pi)^3} 
 \theta_H (\vert \bm{k}_\beta \vert - k_F),
\end{eqnarray} 
where $\theta_H(k)$ is the Heaviside step function and $k_F$ is the Fermi wave number. Figure \ref{fig:diagram} shows the Feynman diagram appropriate for describing exciton creation. No attractive interaction is included in the zeroth order diagrams [\ref{fig:diagram}(a) and \ref{fig:diagram}(b)], while attractive interaction is included in the first order diagrams [\ref{fig:diagram}(${\mathrm c}$) and \ref{fig:diagram}(d)]; this enhances the EH pair density as a result of the exciton creation. Details of this calculation are given in Appendix. 

Sch\"{o}ne and Ekaldt \cite{schone2} studied the screening effect in a jellium model by computing the linear response of the total potential to the sudden creation of a potential induced by a positive charge, with  
\begin{equation}
 \varphi^{\rm ext} (\bm{x};t) = \frac{e}{\vert \bm{x} \vert} \theta (t),
\end{equation}
where $\theta (t)$ is a step function. Within an adiabatic approximation, such a potential can create bound states at $t=0$. The total potential for the single-particle state varies dynamically because of the screening effect and is expressed by the sum of the external and induced potentials
\begin{eqnarray}
 & & \tilde{\varphi}^{\rm tot} (\bm{q};t) \nonumber\\
 &=& \tilde{\varphi}^{\rm ext} (\bm{q};t) + \tilde{\varphi}^{\rm ind} (\bm{q};t), \nonumber\\
 &=& \frac{4\pi e}{\vert \bm{q} \vert^2} \left\{ 1 + \frac{8 e^2}{\hbar\vert \bm{q} \vert^2} 
\int_{0}^{\infty} \frac{d\omega}{\omega} {\rm Im} \tilde{D}_{1}^{\mathrm{R}}(\bm{q};\omega) 
(1-\cos \omega t) \right\} \theta (t), \nonumber\\
\label{eq:pot1}
\end{eqnarray}
where $\tilde{D}_{1}^{\mathrm{R}}(\bm{q};\omega) $ is the retarded density-density correlation function \cite{schone2,fetter_walecka}. Solving the effective-mass equation with the use of Eq.~(\ref{eq:pot1}) yields bound states with time-dependent energy. For large time values, the total potential approaches the Tomas-Fermi potential, $\sim (q_{\rm TF}^{2} + \vert \bm{q}\vert^2)^{-1}$ ($q_{\rm TF}$ is the Tomas-Fermi wave vector), resulting in an absence of bound states. The use of $\tilde{\varphi}^{\rm tot} (\bm{q};\omega)$ in Eq.~(\ref{eq:pot1}) as an external potential together with Eq.~(\ref{eq:n2qq}) enables us to simultaneously study both the attractive interaction effect and the screening effect on the dynamics of the EH pair. We obtain
\begin{eqnarray}
\delta \langle \tilde{n}_2(q,q';t) \rangle 
 &=& - \frac{2 e^2}{\hbar \vert \bm{q} + \bm{q}' \vert^2}  ( I_0 + I_1 ) \theta (t),
 \label{eq:app1}
 \end{eqnarray}
 where
\begin{eqnarray}
 I_0 &=& 4\int_{0}^{\infty} \frac{d\omega' }{\omega'}
 {\rm Im} \tilde{D}_{2}^{\rm R}(q,q',\sigma'';\omega') (1-\cos \omega' t), \label{eq:I_0} \\ 
I_1 &=& \frac{8 e^2}{\hbar\vert \bm{q} + \bm{q}' \vert^2} 
\int_{0}^{\infty} \frac{d\omega_0}{\omega_0} 
{\rm Im} \tilde{D}_{1}^{\mathrm{R}}(\bm{q}+\bm{q}';\omega_0) \nonumber\\
&\times&
\left[
I_0 -
 4 \int_{0}^{\infty} \frac{d\omega'}{\omega'} 
 {\rm Im} \tilde{D}_{2}^{\rm R}(q,q',\sigma'';\omega') C(\omega_0, \omega')
 \right].
 \nonumber\\
 \label{eq:I_1}
\end{eqnarray} 
Here, $C(\omega_0, \omega') = \omega'^{2} (\cos \omega_0 t - \cos \omega' t)/(\omega'^{2}- \omega_{0}^{2})$, $I_0$ represents the EH pair creation, and $I_1$ describes the EH pair annihilation due to the screening. Note that Eq. (\ref{eq:app1}) is regarded as a generalization of the work of Canright \cite{canright}, in which the transient screening response of the electron gas to a suddenly created point charge, $\delta \langle n_1(x;t) \rangle$, is calculated. Given that the stability of the exciton is determined by the previously mentioned competing effect, we define the exciton density as 
\begin{eqnarray}
 \tilde{n}_{\mathrm{exc}}(\bm{q},\bm{q}';t) &=& \sum_{\sigma,\sigma'}  \tilde{n}_{\mathrm{exc}}(q,q';t),
 \label{eq:def} \\
 \tilde{n}_{\mathrm{exc}}(q,q';t) 
 &=& \delta \langle \tilde{n}_2(q,q';t) \rangle 
- \delta \langle \tilde{n}_{2}^{(0)}(q,q';t) \rangle, 
\label{eq:def_a}
\end{eqnarray}
where
\begin{eqnarray}
 \delta \langle \tilde{n}_{2}^{(0)}(q,q';t) \rangle
 = - \frac{2e^2}{\hbar\vert \bm{q} + \bm{q}' \vert^2} I_{0}^{(0)}  \theta (t),
 \label{eq:n20}
\end{eqnarray}
with $I_{0}^{(0)}$ being the zeroth order contribution of Eq.~(\ref{eq:I_0}). Both the attractive interaction between the electron and hole and the screening effect are included in $\delta \langle \tilde{n}_2(q,q';t) \rangle$, while no such effects are included in $\delta \langle \tilde{n}_{2}^{(0)}(q,q';t) \rangle$. Positive and negative values of $ \tilde{n}_{\mathrm{exc}}(\bm{q},\bm{q}';t)$ indicate the existence or absence of the exciton, respectively. If $\tilde{n}_{\mathrm{exc}}(\bm{q},\bm{q}';t)$ has a positive value within a very short time scale, such an exciton can be deemed a {\it transient} exciton.

\section{Application to a jellium model}
One of the main results in this study is a derivation for the exciton density, as shown in Eqs.~(\ref{eq:def}), (\ref{eq:def_a}), and (\ref{eq:n20}). As a trivial example, the application to the exciton in a jellium model without the screening effect is shown because the lifetime of such an exciton is infinite (see Sec.~\ref{verification}). In Sec.~\ref{screening}, the transient nature of the exciton is revealed by considering the screening effect. In Sec.~\ref{higher}, the effect of the higher order perturbation expansion terms is discussed.
\subsection{Without screening effect}
\label{verification}
\begin{figure}[t]
\center
\includegraphics[scale=0.3,clip]{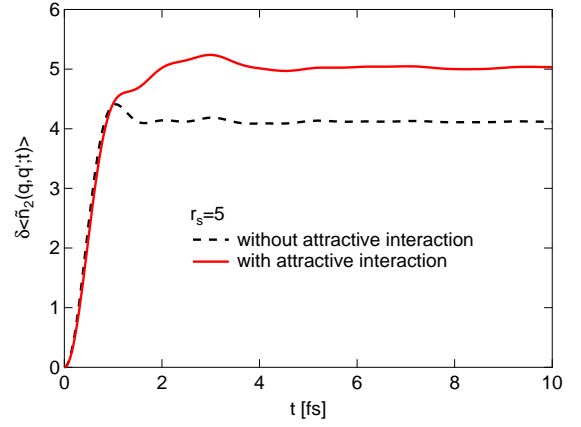}
\caption{\label{fig:wosc}
(Color online) $t$-dependence of $\delta \langle n_{2}(\bm{q},\bm{q}'; t)\rangle$ and $\delta \langle n_{2}^{(0)}(\bm{q},\bm{q}'; t)\rangle$ (see the text for the definition) for $r_s=5$. The parameters are $\vert \bm{q} \vert = \vert \bm{q}' \vert = k_F/2$ and $\vert \bm{q} + \bm{q}' \vert = k_F$.}
\end{figure}

Figure \ref{fig:wosc} shows $t$-dependence of $\delta \langle n_{2}(\bm{q},\bm{q}'; t)\rangle = \sum_{\sigma,\sigma'} \delta \langle n_{2}(q,q'; t)\rangle$ and $\delta \langle n_{2}^{(0)}(\bm{q},\bm{q}'; t)\rangle = \sum_{\sigma,\sigma'} \delta \langle n_{2}^{(0)}(q,q'; t)\rangle$, where the former is the EH pair density with the attractive interaction only while the latter is the EH pair density without both the EH attractive interaction and the screening effect. By definition, the difference between them is the exciton density [see Eq.~(\ref{eq:def_a})]. The density parameter is set to $r_s=5$. The exciton density gradually increases after 1 fs and reaches a positive constant at $t\rightarrow\infty$. Similar behavior is observed for all $r_s$. These results show the validity of the definition for the exciton density given in Eqs.~(\ref{eq:def}), (\ref{eq:def_a}), and (\ref{eq:n20}). 

\begin{figure}[t]
\center
\includegraphics[scale=0.3,clip]{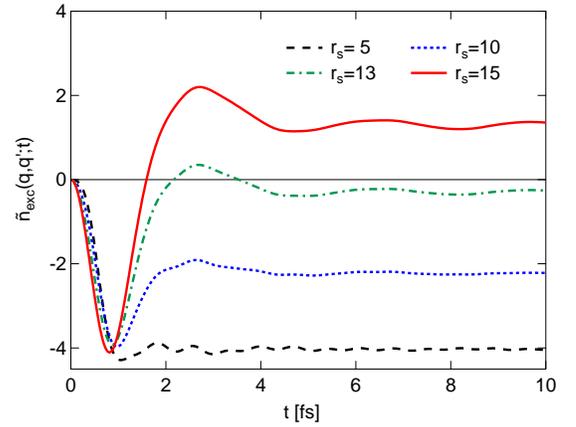}
\caption{\label{fig:rss}
(Color online) $t$-dependence of $\tilde{n}_{\rm exc}(\bm{q},\bm{q}';t)$ for various $r_s$. The parameters are $\vert \bm{q} \vert = \vert \bm{q}' \vert = k_F/2$ and $\vert \bm{q} + \bm{q}' \vert = k_F$. Positive values indicate the existence of the exciton. }
\end{figure}

\begin{figure}[t]
\center
\includegraphics[scale=0.3,clip]{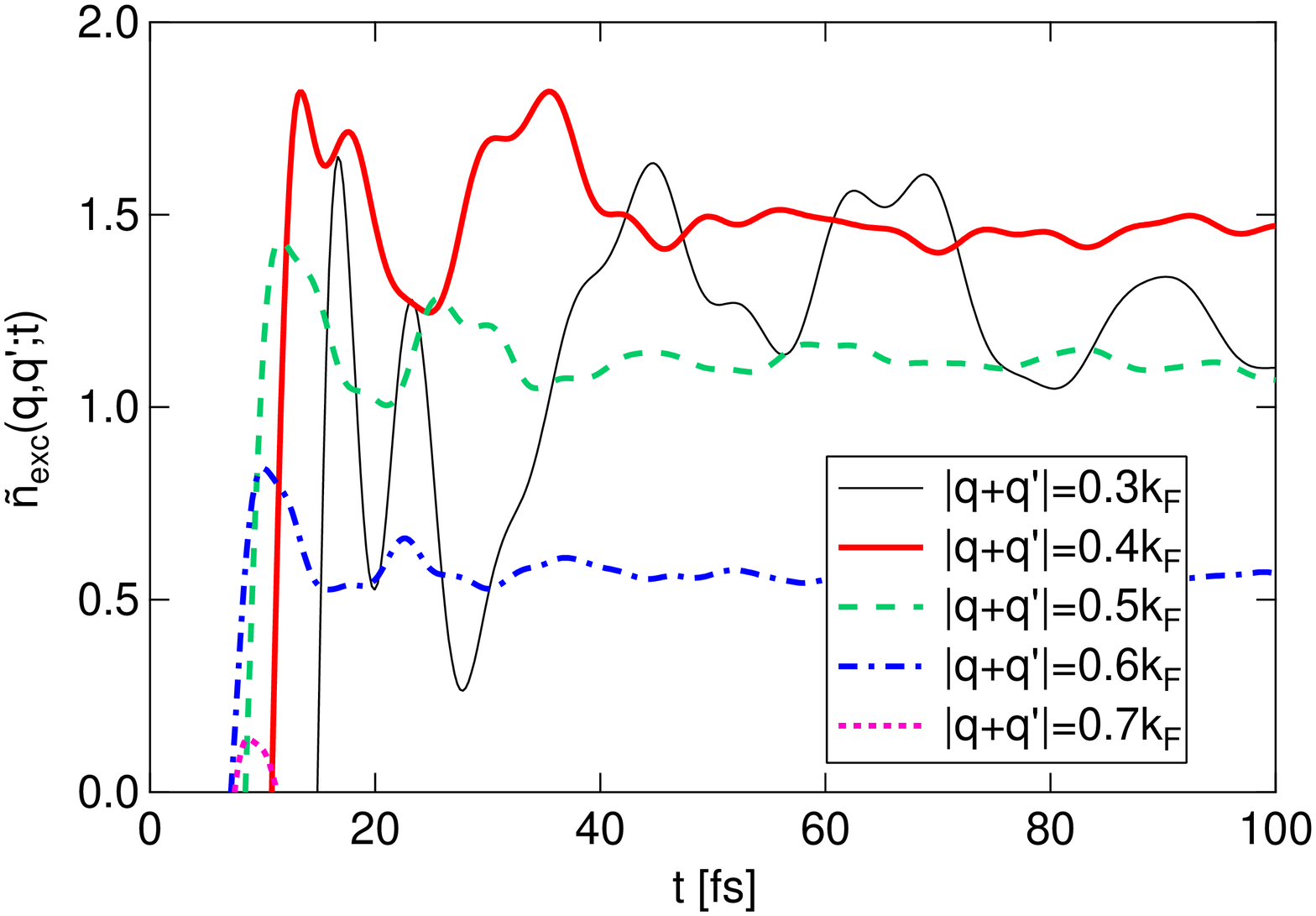}
\caption{\label{fig:q-dep}
(Color online) $t$-dependence of $\tilde{n}_{\rm exc}(\bm{q},\bm{q}';t)$ for various $\vert \bm{q} + \bm{q}' \vert$. The density parameter is set to $r_s = 7$.}
\end{figure}

\subsection{With screening effect}
\label{screening}
The fundamental properties of the transient exciton are investigated by changing the density parameter $r_s$ and the total momentums $\vert \bm{q} + \bm{q}' \vert$, in Secs.~\ref{rsdependence} and \ref{qdependence}, respectively, and the real space analysis is performed in Sec.~\ref{realspace}. Finally, in Sec.~\ref{transientexciton}, the existence of the transient exciton in the jellium model is discussed.
\subsubsection{$r_s$-dependence}
\label{rsdependence}
Figure \ref{fig:rss} shows the exciton density in a jellium model calculated using Eq.~(\ref{eq:def}) for $r_s=5,10,13$, and $15$. We set $\vert \bm{q}\vert = \vert \bm{q}' \vert = \vert \bm{q}+\bm{q}' \vert/2 =k_F/2$, i.e., the total momentum of the exciton is $k_F$, while the relative momentum is zero. The screening effect in $\tilde{D}_{1}^{\rm R}(\bm{q};\omega')$ is treated within the random phase approximation (RPA) \cite{fetter_walecka}, together with the local field correction (LFC). The analytic expression for the dielectric screening function given in Ref.~\cite{ichimaru} is used for the LFC \cite{note1}. We compute the integral involving $\tilde{D}_{2}^{\rm R}(q,q',\sigma'';\omega')$ using the standard Monte Carlo approach. When $t$ is less than 1 fs, $\tilde{n}_{\rm exc}(\bm{q},\bm{q}';t)$ decreases for all $r_s$. When $t$ is larger than 1 fs, $\tilde{n}_{\rm exc}(\bm{q},\bm{q}';t)$ begins to increase for large $r_s$. In the case of $r_s=13$, $\tilde{n}_{\rm exc}(\bm{q},\bm{q}';t)$ is positive only when $t=2$--$4$ fs, which can be interpreted as indicating a transient exciton. In the case of $r_s=15$, on the other hand, $\tilde{n}_{\rm exc}(\bm{q},\bm{q}';t)$ is positive when $t>1.8$ fs, which is interpreted as evidence of a stable exciton \cite{note2}. This clearly shows a crossover from a transient to a stable exciton, which is due to the weak screening effect that occurs in low-density electron gas systems. 

\begin{figure}[t]
\center
\includegraphics[scale=0.35,clip]{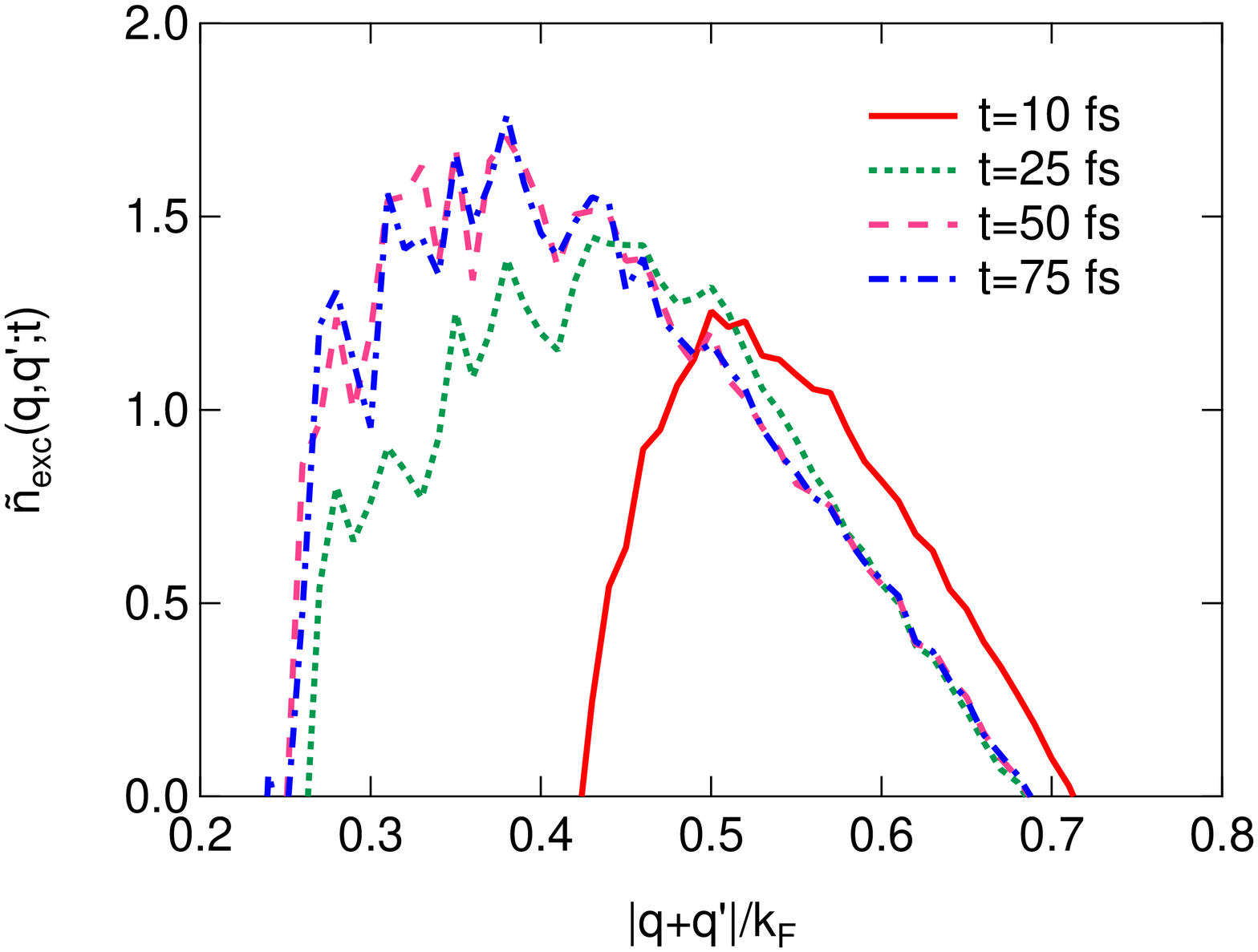}
\caption{\label{fig:q-spectra}
(Color online) $\vert \bm{q}+\bm{q}' \vert$-dependence of $\tilde{n}_{\rm exc}(\bm{q},\bm{q}';t)$ for $t=10,25,50$, and $75$ fs.}
\end{figure}

\begin{figure}[t]
\center
\includegraphics[scale=0.35,clip]{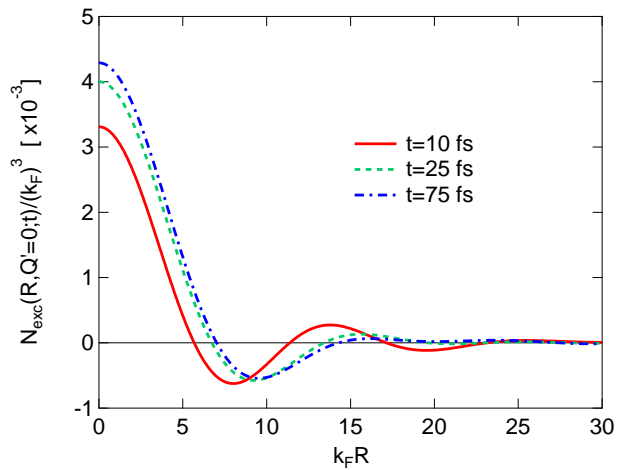}
\caption{\label{fig:R-spectra}
(Color online) Real-space distribution of $\tilde{N}_{\rm exc}(R;\bm{Q}'=\bm{0};t)$ for $t=10,25$, and $75$ fs. The spatial period of the exciton density oscillation increases with time.}
\end{figure}

\subsubsection{$q$-dependence}
\label{qdependence}
Figure \ref{fig:q-dep} shows the $t$-dependence of $\tilde{n}_{\rm exc}(\bm{q},\bm{q}';t)$ in the case of $r_s=7$ for various $\vert \bm{q}+\bm{q}' \vert$. In the initial stage ($t\sim 10$ fs), large $\vert \bm{q}+\bm{q}' \vert/k_F =0.5$--$0.7$ contributes to exciton formation while, in the final stage ($t>20$ fs), small $\vert \bm{q}+\bm{q}' \vert/k_F =0.3$--$0.5$ also contributes to the formation. This means that both the average radius and the spatial period of the exciton density oscillation in real space increase with time, which is a novel property involving the exciton formation. The Fourier analysis of the density in space and time is given below. 

\subsubsection{Real space analysis}
\label{realspace}
Figure \ref{fig:q-spectra} shows a snapshot of $\vert \bm{q}+\bm{q}' \vert$-dependence of $\tilde{n}_{\rm exc}(\bm{q},\bm{q}';t)$ at $t=10, 25, 50$, and $75$ fs. At $t=10$ fs, $\tilde{n}_{\rm exc}(\bm{q},\bm{q}';t)$ is positive when $0.42k_F \le \vert \bm{q}+\bm{q}' \vert \le 0.71k_F$ and takes the maximum at $\vert \bm{q}+\bm{q}' \vert\simeq0.5k_F$. As $t$ increases, the value of $\tilde{n}_{\rm exc}(\bm{q},\bm{q}';t)$ having small $\vert \bm{q}+\bm{q}' \vert$ increases, as mentioned in the main text: for example, at $t=75$ fs, the peak of $\tilde{n}_{\rm exc}(\bm{q},\bm{q}';t)$ redshifts and the width of $\tilde{n}_{\rm exc}(\bm{q},\bm{q}';t)$ increases ($0.25k_F \le \vert \bm{q}+\bm{q}' \vert \le 0.69k_F$). To study a real space distribution of excitons, we use the Fourier transformation given by
\begin{eqnarray}
 n_{\rm exc}(\bm{x},\bm{x}';t) 
 &=&
    \int \frac{d\bm{q}}{(2\pi)^3}
    \int \frac{d\bm{q}'}{(2\pi)^3} \nonumber\\
  &\times&
    e^{i \bm{q} \cdot (\bm{x} - \bm{x}'') } 
    e^{i \bm{q}' \cdot (\bm{x}' - \bm{x}'') }
    \tilde{n}_{\rm exc}(\bm{q},\bm{q}';t).
    \nonumber\\
\end{eqnarray}
Let $\bm{Q}$ and $\bm{Q}'$ be the total momentum and relative momentum of a exciton, respectively, i.e.,
\begin{eqnarray}
    \bm{Q} = \bm{q} + \bm{q}', \ \ \
    \bm{Q}' = \frac{\bm{q} - \bm{q}'}{2}.
\end{eqnarray}
Then, the exciton density can be represented by
\begin{eqnarray}
& & n_{\rm exc}(\bm{x},\bm{x}';t) \nonumber\\
 &=&
    \int \frac{d\bm{Q}'}{(2\pi)^3} 
    e^{i \bm{Q}' \cdot (\bm{x} - \bm{x}') } \nonumber\\
    &\times&
    \left[
    \int \frac{d\bm{Q}}{(2\pi)^3}
    e^{i \bm{Q} \cdot \bm{R} } 
    \tilde{n}_{\rm exc}\left(\frac{\bm{Q}}{2}+\bm{Q}',\frac{\bm{Q}}{2}-\bm{Q}';t\right)  \right] \nonumber\\
 &\equiv&
    \int \frac{d\bm{Q}'}{(2\pi)^3} 
    e^{i \bm{Q}' \cdot (\bm{x} - \bm{x}') }
    \tilde{N}_{\rm exc}(\bm{R};\bm{Q}';t),
\end{eqnarray}
where $\bm{R} = (\bm{x} + \bm{x}' - 2\bm{x}'')/2$ is the position of the center of mass and $\tilde{N}_{\rm exc}(\bm{R};\bm{Q}';t)$ is the exciton density in mixed coordinates $\bm{R}$ and $\bm{Q}'$. In this work, we study $\tilde{N}_{\rm exc}(\bm{R};\bm{Q}'=\bm{0};t)$ only. The case of $\bm{Q}'\ne\bm{0}$ will be studied elsewhere \cite{note3}. In a jellium model, the exciton density depends on $R=\vert \bm{R} \vert$ and is written as
\begin{eqnarray}
 \tilde{N}_{\rm exc}(R;\bm{Q}'=\bm{0};t)
 &=&  \int \frac{d\bm{Q}}{(2\pi)^3}
    e^{i \bm{Q} \cdot \bm{R} } 
    \tilde{n}_{\rm exc}\left(\frac{\bm{Q}}{2},\frac{\bm{Q}}{2};t\right) 
    \nonumber\\
  &=& \frac{1}{2\pi^2 R} \int_{0}^{\infty} dQ Q
  \sin (Q R) \tilde{n}_{\rm exc} (Q;t).
  \label{eq:nexc}
  \nonumber\\
\end{eqnarray}
Figure~\ref{fig:R-spectra} shows the distribution of $\tilde{N}_{\rm exc}(R;\bm{Q}'=\bm{0};t)$ given by Eq.~(\ref{eq:nexc}) for $t=10, 25$, and $75$ fs. At $t=10$ fs, the exciton exists at the region $k_FR\le 5.6$, $11.4\le k_FR \le 17.1$, and $23.1\le k_FR \le 30$, although the magnitude of $\tilde{N}_{\rm exc}(R;\bm{Q}'=\bm{0};t)$ decreases drastically as $k_FR$ increases. As $t$ increases, the magnitude of $\tilde{N}_{\rm exc}(R;\bm{Q}'=\bm{0};t)$ near $R\simeq 0$ increases and the exciton exists at the region $k_FR\le 7$ and $14.3\le k_FR \le 27.5$. This means that both the average radius of the exciton and the spatial period of the density oscillation in $\tilde{N}_{\rm exc}(R;\bm{Q}'=\bm{0};t)$ increases with time. This is due to an increase in the value of $\tilde{n}_{\rm exc}(\bm{q},\bm{q}';t)$ with small $\vert \bm{q}+\bm{q}' \vert$, as shown in Fig.~\ref{fig:q-spectra}. We expect that the character of the exciton time evolution in a realistic material is qualitatively the same as that in a jellium model.

\begin{figure}[t]
\center
\includegraphics[scale=0.40,clip]{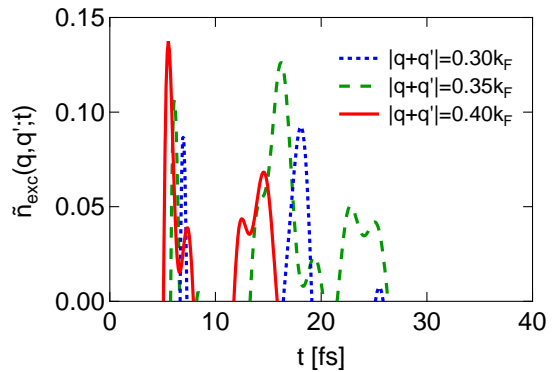}
\caption{\label{fig:transient}
(Color online) $t$-dependence of $\tilde{n}_{\rm exc}(\bm{q},\bm{q}';t)$ for various $\vert \bm{q} + \bm{q}' \vert$. The density parameter is set to $r_s=$4.5.}
\end{figure}

\subsubsection{Transient exciton regime}
\label{transientexciton}
As shown in Fig.~\ref{fig:q-dep} (for the case of $r_s=7$), the stable exciton exists even at $t\rightarrow \infty$: the exciton with $\vert \bm{q}+\bm{q}' \vert = 0.7k_F$ is the transient exciton, whereas those with $\vert \bm{q}+\bm{q}' \vert = 0.3k_F$ to $0.6k_F$ are the stable exciton. By considering the parameter range $\vert \bm{q}+\bm{q}' \vert =$0.1$k_F$--1.0$k_F$, the specific $r_s$ in which the transient exciton exists only has been investigated. Figure \ref{fig:transient} shows $t$-dependence of $\tilde{n}_{\rm exc}(\bm{q},\bm{q}';t)$ for $\vert \bm{q} + \bm{q}' \vert /k_F=0.3, 0.35$, and 0.4. $r_s=4.5$ was used. The transient exciton is clearly observed (positive $\tilde{n}_{\rm exc}$), i.e., the exciton density vanishes at $t\rightarrow \infty$. No exciton modes are observed for other $\vert \bm{q} + \bm{q}' \vert$s (negative $\tilde{n}_{\rm exc}$). Thus, for the specific density $r_s=4.5$, the sudden creation of a positive charge into a jellium model creates the transient exciton only. This is quite reasonable because the most real metals have $r_s=$2--5. Through the thorough investigation, with the density parameter $r_s$ below 4.0 and above 5.0, no exciton modes and stable exciton modes were observed, respectively. This is also physically reasonable because in such a system having a high or low electron density, the screening becomes complete or incomplete enough to generate no excitons or stable excitons, respectively. 


The transient exciton has been observed at silver surface by Cui et al. \cite{Cui}. The present calculation suggests that the transient exciton can exist in a small range around $r_s=$4.5. Based on this result, it could be predicted that the stable exciton should be observed if the electron density is decreased at the silver surface. The chemical adsorption at the surface (such as oxygen adsorption) may be useful to examine the crossover from the transient to stable exciton.

It should be noted that on the femtosecond timescale, the existence of the transient exciton is limited by the uncertainty relation between time and energy, i.e., $\Delta t \Delta E \ge \hbar/2$: for example, when $\Delta t\sim 1$ fs, we obtain $\Delta E\sim0.33$ eV. With such a large uncertainty, detemining the energy of the transient exciton would be meaningless.

\subsection{Higher order perturbation expansion}
\label{higher}
\begin{figure}[t]
\center
\includegraphics[scale=0.3,clip]{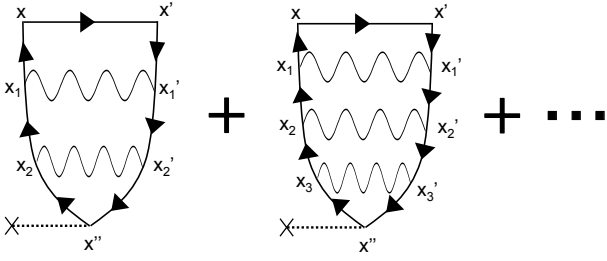}
\caption{\label{fig:diagram2} The second and third order contribution to $D_{2}^{\rm T}(x,x',x'';t-t')$. }
\end{figure}

In the $D_{2}^{\rm T}(x,x',x'';t-t')$ computation, one may find $(2m+3)!$ possible diagrams in the $m$th order ($m$ is a non-negative integer) by applying Wick's theorem \cite{fetter_walecka}. The present calculation considers up to the first order contribution of $D_{2}^{\rm T}(x,x',x'';t-t')$ (shown in Figs.~\ref{fig:diagram}(${\mathrm c}$) and \ref{fig:diagram}(d)), which are the most fundamental components as regards examination of exciton creation. Note that the inclusion of higher order terms (such as Fig.~\ref{fig:diagram2}) may enhance the magnitude of $\tilde{n}_{\rm exc}(\bm{q},\bm{q}';t)$. The inclusion of infinite terms leads to nonperturbative treatment, which is desirable for the complete description of the exciton. However, it is quite difficult to perform such a calculation at the present formulation. A new approach for the nonperturbative treatment has to be developed.

\section{Summary}
The purpose of this paper was to develop a theory to describe the dynamics of excitons with time evolutions that cannot be studied using the effective-mass equation and the standard BSE with stationary interaction potential. The linear response of the EH pair density to an external perturbation was examined. By considering the electron-hole attractive interaction and the screening effect, a definition of exciton density in a first-principles context was derived. Further, the application of the proposed theory to a jellium model confirmed the existence of transient excitons and unveiled the mechanism of exciton formation. Investigating the associated band structure, spin, phonon, and quantum size effects will provide an enhanced understanding of the dynamics of excitons in condensed matter systems.

\section{Acknowledgements}
The author acknowledges S. G. Louie, M. Bernadi, J. Mustafa, and K. Ohno for their contributions to fruitful discussions. This research was supported by the CMRI (Computational Materials Research Initiative, HPCI MEXT) program for Young Researcher Overseas Visits and a grant-in-aid for Scientific Research in Innovative Areas (Grant No. 25104713) from MEXT.  

\appendix
\section{Zeroth order contribution}
\label{app1}
First, we calculate the zeroth order contributions in $\tilde{D}_{2}^{\mathrm{T}} (q,q',\sigma'';\omega)$ via the Fourier transformations. By using the interaction picture [$\psi_I (x;t)=e^{iH_0t/\hbar} \psi(x) e^{-iH_0t/\hbar}$ (the subscript $I$ is omitted for the simplicity.)] and the Wick's theorem, the following time-ordered product
\begin{widetext}
\begin{eqnarray}
  \langle N,0 \vert {\mathrm T} [ \psi^\dagger(x';t) \psi(x;t) \psi^\dagger(x;t) \psi(x';t) \psi^\dagger(x'';t')\psi(x'';t')] \vert N, 0 \rangle, 
\end{eqnarray}
can be decomposed into six terms
\begin{eqnarray} 
&+& \left[   \psi^\dagger(x';t) \psi(x;t) \right]_c  \left[ \psi^\dagger(x;t) \psi(x';t)  \right]_c   \left[   \psi^\dagger(x'';t')\psi(x'';t') \right]_c 
\nonumber\\
&-&
\left[   \psi^\dagger(x';t) \psi(x;t) \right]_c  \left[ \psi^\dagger(x;t) \psi(x'';t')  \right]_c   \left[   \psi^\dagger(x'';t')\psi(x';t) \right]_c
\nonumber\\
&-&
\left[   \psi^\dagger(x';t) \psi(x';t) \right]_c  \left[ \psi^\dagger(x;t) \psi(x;t)  \right]_c   \left[   \psi^\dagger(x'';t')\psi(x'';t') \right]_c
\nonumber\\
&+&
\left[   \psi^\dagger(x';t) \psi(x';t) \right]_c  \left[ \psi^\dagger(x;t) \psi(x'';t')  \right]_c   \left[   \psi^\dagger(x'';t')\psi(x;t) \right]_c
\nonumber\\
&+&
\left[   \psi^\dagger(x';t) \psi(x'';t') \right]_c  \left[ \psi^\dagger(x;t) \psi(x;t)  \right]_c   \left[   \psi^\dagger(x'';t')\psi(x';t) \right]_c
\nonumber\\
&-&
\left[   \psi^\dagger(x';t) \psi(x'';t') \right]_c  \left[ \psi^\dagger(x;t) \psi(x';t)  \right]_c   \left[   \psi^\dagger(x'';t')\psi(x;t) \right]_c,
\label{eq:ppc}
\end{eqnarray}
\end{widetext}
where $[\cdots]_c$ denotes the contraction (see Ref.~[14]), which is expressed by the non-interacting Green's function via the relations
\begin{eqnarray}
 \left[   \psi^\dagger(x';t)\psi(x'';t') \right]_c &=& -i G^{(0)} (x'', x'; t'- t), \\
 \left[   \psi^\dagger(x;t)\psi(x';t) \right]_c &=& -i G^{(0)} (x', x; t - t^+).  
\end{eqnarray} 
The Fourier transformation in time of Eq.~(\ref{eq:ppc}) multiplied by $(-i)$ yields the zeroth order contributions $D_{2}^{\mathrm{T}(0)}(x,x',x'';\omega)$. The first and third terms in Eq.~(\ref{eq:ppc}) give ${\rm Im} D_{2}^{\mathrm{T}(0)}(x,x',x'';\omega) \propto \delta (\omega)$ that does not contribute $\delta \langle \tilde{n}_2(q,q';t) \rangle$ because $(1-\cos \omega t)/\omega \rightarrow 0$ [see Eqs.~(\ref{eq:I_0}) and (\ref{eq:I_1})]. Thus, the zeroth order contribution $D_{2}^{\mathrm{T}(0)}(x,x',x'';\omega)$ is expressed by the sum of four terms
\begin{equation}
 D_{2}^{\mathrm{T}(0)}(x,x',x'';\omega) 
 = \sum_{i=1}^{4} D_{2}^{\mathrm{T}(0-i)}(x,x',x'';\omega), 
\end{equation}
where
\begin{widetext}
\begin{eqnarray*}
 D_{2}^{\mathrm{T}(0-1)}(x,x',x'';\omega) 
 &=&
   \delta_{\sigma\sigma'} \delta_{\sigma''\sigma} \delta_{\sigma'\sigma''} 
 (-1) \left( i \sum_{\alpha}^{\rm occ}
 \phi_{\alpha\sigma} (\bm{x}) \phi_{\alpha\sigma}^{*} (\bm{x}') \right)
 \int \frac{d\omega'}{2\pi} 
 G_{\sigma''}^{(0)} (\bm{x}'',\bm{x}; \omega') 
 G_{\sigma'}^{(0)} (\bm{x}',\bm{x}'';\omega + \omega'),
 \nonumber\\
 D_{2}^{\mathrm{T}(0-2)}(x,x',x'';\omega) 
 &=&
  \delta_{\sigma'\sigma'} \delta_{\sigma''\sigma} \delta_{\sigma\sigma''} 
  \left( i \sum_{\alpha}^{\rm occ}
  \vert \phi_{\alpha\sigma'} (\bm{x}') \vert^2 \right)
 \int \frac{d\omega'}{2\pi} 
 G_{\sigma''}^{(0)} (\bm{x}'',\bm{x}; \omega') 
 G_{\sigma}^{(0)} (\bm{x},\bm{x}'';\omega + \omega'),
 \nonumber\\
  D_{2}^{\mathrm{T}(0-3)}(x,x',x'';\omega) 
 &=&
    \delta_{\sigma\sigma} \delta_{\sigma''\sigma'} \delta_{\sigma'\sigma''} 
  \left( i \sum_{\alpha}^{\rm occ}
  \vert \phi_{\alpha\sigma} (\bm{x}) \vert^2 \right)
 \int \frac{d\omega'}{2\pi}  
 G_{\sigma''}^{(0)} (\bm{x}'',\bm{x}'; \omega')
 G_{\sigma'}^{(0)} (\bm{x}',\bm{x}'';\omega + \omega'),
 \nonumber\\
  D_{2}^{\mathrm{T}(0-4)}(x,x',x'';\omega) 
 &=&
   \delta_{\sigma'\sigma} \delta_{\sigma''\sigma'} \delta_{\sigma\sigma''} 
 (-1) \left( i \sum_{\alpha}^{\rm occ}
  \phi_{\alpha\sigma'} (\bm{x}') \phi_{\alpha\sigma'}^{*} (\bm{x}) \right)
 \int \frac{d\omega'}{2\pi}  
 G_{\sigma''}^{(0)} (\bm{x}'',\bm{x}'; \omega') 
 G_{\sigma}^{(0)} (\bm{x},\bm{x}'';\omega + \omega').
\end{eqnarray*}
\end{widetext}
The integration for $\omega'$ can be evaluated by using the expression
\begin{eqnarray}
  & & \int \frac{d\omega'}{2\pi}  
  G_{\sigma'}^{(0)}(\bm{x}',\bm{x}'';\omega') 
  G_{\sigma}^{(0)}(\bm{x},\bm{x}''';\omega + \omega')
  \nonumber\\
 &=& 
 i \sum_{c}^{{\rm emp}} \sum_{v}^{{\rm occ}} \Big[
 \frac{ \phi_{c\sigma}(\bm{x}) \phi_{v\sigma'}(\bm{x}') 
           \phi_{v\sigma'}^{*}(\bm{x}'') \phi_{c\sigma}^{*}(\bm{x}''')}
        { \omega - (\omega_{\bm{k}_c} - \omega_{\bm{k}_v}) + i\delta } 
        \nonumber\\
&-&  \frac{ \phi_{v\sigma}(\bm{x}) \phi_{c\sigma'}(\bm{x}')
         \phi_{c\sigma'}^{*}(\bm{x}'') \phi_{v\sigma}^{*}(\bm{x}''')}
        { \omega - (\omega_{\bm{k}_v} - \omega_{\bm{k}_c}) - i\delta } \Big].
        \label{eq:L01234}
\end{eqnarray}
The imaginary part of the correlation function is calculated by using the identity $(\omega \pm i\delta)^{-1} = {\cal P}\omega^{-1} \mp i\pi \delta (\omega)$ valid for real $\omega$. In the following, we calculate these four terms relevant to an increase in the EH pair density.
 
\subsection{First term}
The contribution from $D_{2}^{\mathrm{T}(0-1)}(x,x',x'';\omega)$ corresponds to the that from Fig.~\ref{fig:diagram}(b). By using Eq.~(\ref{eq:pw}), $D_{2}^{\mathrm{T}(0-1)}(x,x',x'';\omega)$ is written as 
\begin{widetext}
\begin{eqnarray}
 D_{2}^{\mathrm{T}(0-1)}(x,x',x'';\omega) 
 &=&
    \delta_{\sigma\sigma'} \delta_{\sigma''\sigma} \delta_{\sigma'\sigma''} 
    \int \frac{d\bm{k}_\alpha}{(2\pi)^3}
    \int \frac{d\bm{k}_c}{(2\pi)^3}
    \int \frac{d\bm{k}_v}{(2\pi)^3}  
       \theta_H (k_F - \vert \bm{k}_\alpha \vert)
       \theta_H (\vert \bm{k}_c \vert - k_F)
       \theta_H (k_F - \vert \bm{k}_v \vert) \nonumber\\
& & \times
 \left[
 \frac{  e^{i ( \bm{k}_\alpha - \bm{k}_v ) \cdot (\bm{x} - \bm{x}'') } 
           e^{i ( \bm{k}_c - \bm{k}_\alpha ) \cdot (\bm{x}' - \bm{x}'') }   }
        { \omega - (\omega_{\bm{k}_c} - \omega_{\bm{k}_v}) + i\delta }
-  \frac{  e^{i ( \bm{k}_\alpha - \bm{k}_c ) \cdot (\bm{x} - \bm{x}'') } 
              e^{i ( \bm{k}_v - \bm{k}_\alpha ) \cdot (\bm{x}' - \bm{x}'') }    }
        { \omega - (\omega_{\bm{k}_v} - \omega_{\bm{k}_c}) - i\delta } \right].
\end{eqnarray}
The Fourier transformation in space of $D_{2}^{\mathrm{T}(0-1)}(x,x',x'';\omega)$ yields 
\begin{eqnarray}
 \tilde{D}_{2}^{\mathrm{T}(0-1)}(q,q',\sigma'';\omega)
 &=& \delta_{\sigma\sigma'} \delta_{\sigma''\sigma} \delta_{\sigma'\sigma''}
    \int \frac{d\bm{k}_v}{(2\pi)^3}  \theta (k_F - \vert \bm{k}_v \vert) \nonumber\\
& & \times
 \left[
 \frac{ 
 \theta_H (\vert \bm{k}_v + \bm{q} + \bm{q}' \vert - k_F) 
 \theta_H (k_F - \vert \bm{k}_v + \bm{q} \vert) } 
      { \omega - (\omega_{\bm{k}_v + \bm{q} + \bm{q}'} - \omega_{\bm{k}_v}) + i\delta }
-  
\frac{ \theta_H (\vert \bm{k}_v- \bm{q} - \bm{q}' \vert - k_F) \theta_H (k_F - \vert \bm{k}_v - \bm{q}' \vert) } 
        { \omega - (\omega_{\bm{k}_v} - \omega_{\bm{k}_v - \bm{q} - \bm{q}'}) - i\delta } \right], 
         \nonumber\\
\end{eqnarray}
whose imaginary part is given as
\begin{eqnarray}
{\rm Im}  \tilde{D}_{2}^{\mathrm{T}(0-1)}(q,q',\sigma'';\omega) 
 &=&
    \delta_{\sigma\sigma'} \delta_{\sigma''\sigma} \delta_{\sigma'\sigma''} (- \pi) 
    \int \frac{d\bm{k}_v}{(2\pi)^3}  
       \theta (k_F - \vert \bm{k}_v \vert) \nonumber\\
& & \times
 \big[
 \theta_H (\vert \bm{k}_v + \bm{q} + \bm{q}' \vert - k_F) 
 \theta_H (k_F - \vert \bm{k}_v + \bm{q} \vert)
        \delta ( \omega - \omega_{\bm{k}_v + \bm{q} + \bm{q}'} + \omega_{\bm{k}_v} ) \nonumber\\
& & + \theta_H (\vert \bm{k}_v- \bm{q} - \bm{q}' \vert - k_F)
        \theta_H (k_F - \vert \bm{k}_v - \bm{q}' \vert)
       \delta ( \omega - \omega_{\bm{k}_v} + \omega_{\bm{k}_v - \bm{q} - \bm{q}'} ) \big]. 
\end{eqnarray}
Since $\omega_{\bm{k}_c} - \omega_{\bm{k}_v} > 0$ and $\omega > 0$, the second term in the square bracket of this equation vanishes. Thus, we obtain 
\begin{eqnarray}
{\rm Im} \tilde{D}_{2}^{\mathrm{T}(0-1)}(q,q',\sigma'';\omega)
 &=&
    \delta_{\sigma\sigma'} \delta_{\sigma''\sigma} \delta_{\sigma'\sigma''} (- \pi) 
    \int \frac{d\bm{k}_v}{(2\pi)^3}  
       \theta_H (k_F - \vert \bm{k}_v \vert) 
 \theta_H (\vert \bm{k}_v + \bm{q} + \bm{q}' \vert - k_F) 
 \theta_H (k_F - \vert \bm{k}_v + \bm{q} \vert) \nonumber\\
 & & \times    \delta \left( \omega - \frac{\hbar \vert \bm{k}_v + \bm{q} + \bm{q}' \vert^{2}}{2m} + \frac{\hbar k_{v}^{2}}{2m} \right).   
\end{eqnarray}
\end{widetext}
${\rm Im} \tilde{D}_{2}^{\mathrm{T}(0-1)}(q,q',\sigma'';\omega)$ has a negative sign and depends on the magnitude of vectors $\bm{q} + \bm{q}'$ and $\bm{q}$ the angle $\theta$ between these two vectors.
\subsection{Second and third terms}
Similarly to the derivation of $\tilde{D}_{2}^{\mathrm{T}(0-1)}(q,q',\sigma'';\omega)$, we obtain the imaginary part of $\tilde{D}_{2}^{\mathrm{T}(0-2)}(q,q',\sigma'';\omega)$ for $\omega>0$,
\begin{eqnarray}
& & {\rm Im}\tilde{D}_{2}^{\mathrm{T}(0-2)}(q,q',\sigma'';\omega) \nonumber\\
   &=& 
  \delta_{\sigma'\sigma'} \delta_{\sigma''\sigma} \delta_{\sigma\sigma''}  
  \frac{mk_{F}^{3}\delta ( \bm{q}' )}{6\pi\hbar} \nonumber\\
  &\times&
    \int d\bm{k}
     \theta_H (\vert \bm{k} + \bm{q} \vert - k_F) \theta_H (k_F - \vert \bm{k} \vert) 
     \nonumber\\
     &\times&
     \delta \left( \bm{k} \cdot \bm{q} + \frac{\vert \bm{q}\vert^2}{2} -\frac{m\omega}{\hbar} \right).
\end{eqnarray}
The calculation of the integral for $\bm{k}$ is the same as that of the non-interacting polarization function (see Ref.~[14]). The third term $\tilde{D}_{2}^{\mathrm{T}(0-3)}(q,q',\sigma'';\omega)$ is obtained by transforming $\bm{q} \leftrightarrow \bm{q}'$ for the expression in $\tilde{D}_{2}^{\mathrm{T}(0-2)}(q,q',\sigma'';\omega)$. If we assume that the total momentum of the exciton is not zero (i.e., $\vert \bm{q} + \bm{q}' \vert \ne 0$) and the electron and hole move along the same direction (i.e., $\bm{q}//\bm{q}'$), $\bm{q}$ and $\bm{q}' $ are not equal to zero. In this assumption, these terms, $\tilde{D}_{2}^{\mathrm{T}(0-2)}(q,q',\sigma'';\omega)$ and $\tilde{D}_{2}^{\mathrm{T}(0-3)}(q,q',\sigma'';\omega)$, do not contribute to an increase in the EH pair density due to the presence of the factors $\delta(\bm{q}')$ and $\delta(\bm{q})$.

\subsection{Forth term}
The contribution from $D_{2}^{\mathrm{T}(0-4)}(x,x',x'';\omega)$ corresponds to the that from Fig.~\ref{fig:diagram}(a). This term can be obtained by transforming $\bm{q} \leftrightarrow \bm{q}'$ for the expression in $\tilde{D}_{2}^{\mathrm{T}(0-1)}(q,q',\sigma'';\omega)$. Thus, the imaginary part of $\tilde{D}_{2}^{\mathrm{T}(0-4)}(q,q',\sigma'';\omega)$ for $\omega > 0$ is given as
\begin{widetext}
\begin{eqnarray}
{\rm Im} \tilde{D}_{2}^{\mathrm{T}(0-4)}(q,q',\sigma'';\omega)
 &=&
  \delta_{\sigma'\sigma} \delta_{\sigma''\sigma'} \delta_{\sigma\sigma''}
  ( -\pi) 
    \int \frac{d\bm{k}_v}{(2\pi)^3}  
       \theta_H (k_F - \vert \bm{k}_v \vert) 
 \theta_H (\vert \bm{k}_v + \bm{q} + \bm{q}' \vert - k_F) 
 \theta_H (k_F - \vert \bm{k}_v + \bm{q}' \vert) \nonumber\\
 & & \times   
  \delta \left( \omega - \frac{\hbar \vert \bm{k}_v + \bm{q} + \bm{q}' \vert^{2}}{2m} + \frac{\hbar k_{v}^{2}}{2m} \right).   
 \label{eq:D04}
\end{eqnarray}
\end{widetext}
${\rm Im} \tilde{D}_{2}^{\mathrm{T}(0-4)}(q,q',\sigma'';\omega)$ has a negative sign and depends on the magnitude of vectors $\bm{q} + \bm{q}'$ and $\bm{q}'$ the angle $\theta$ between these two vectors.

\subsection{Formulae for the integral for the zeroth order terms}
The integral appeared in ${\rm Im} \tilde{D}_{2}^{\mathrm{T}(0-1)}(q,q',\sigma'';\omega)$ and ${\rm Im} \tilde{D}_{2}^{\mathrm{T}(0-4)}(q,q',\sigma'';\omega)$ can be calculated analytically. Now we focus on the computation of ${\rm Im} \tilde{D}_{2}^{\mathrm{T}(0-4)}(q,q',\sigma'';\omega)$. The result for ${\rm Im} \tilde{D}_{2}^{\mathrm{T}(0-1)}(q,q',\sigma'';\omega)$ will be obtained by replacing $\bm{q} \leftrightarrow \bm{q}'$ in that for ${\rm Im} \tilde{D}_{2}^{\mathrm{T}(0-4)}(q,q',\sigma'';\omega)$ shown below. 

The method for calculating the non-interacting polarization function (see Ref.~[14]) is useful for performing the integral of $\bm{k}_v$ in Eq.~(\ref{eq:D04}). The $\delta$-function in Eq.~(\ref{eq:D04}) is modified into 
\begin{eqnarray}
& & \delta \left( \omega - \frac{\hbar \vert \bm{k}_v + \bm{q} + \bm{q}' \vert^{2}}{2m} 
 + \frac{\hbar k_{v}^{2}}{2m} \right) \nonumber\\
 &=& 
 \frac{m}{\hbar}
 \delta \left( \bm{k}_v\cdot (\bm{q} + \bm{q}') 
 + \frac{ \vert \bm{q} + \bm{q}' \vert^{2}}{2} - \frac{m\omega}{\hbar} 
  \right).
\end{eqnarray}
Since the vector $\bm{k}_v$ satisfying the equation 
\begin{eqnarray}
 \bm{k}_v \cdot (\bm{q}+\bm{q}') + \frac{\vert \bm{q} + \bm{q}' \vert^2}{2} -\frac{m\omega}{\hbar} 
 &=& (\bm{k}_v - \bm{k}_0) \cdot (\bm{q} + \bm{q}') \nonumber\\
 &=& 0,
\end{eqnarray}
represents the plane perpendicular to the vector $(\bm{q} + \bm{q}')$, the integral for $\bm{k}_v$ represents the area of the intersection of a part of the Fermi sphere with the plane $(\bm{k}_v - \bm{k}_0) \cdot (\bm{q} + \bm{q}') = 0$, where $ \bm{k}_0$ is given by
\begin{equation}
 \bm{k}_0 = z_0 \frac{\bm{q}+ \bm{q}'}{\vert \bm{q} + \bm{q}' \vert}, \ \ 
 z_0 = \frac{m\omega}{\hbar \vert \bm{q}+ \bm{q}'\vert} - \frac{\vert \bm{q}+ \bm{q}'\vert}{2}.
\end{equation}
The part of the Fermi sphere that contributes the integral is determined by the Heaviside step functions $\theta_H$ in Eq.~(\ref{eq:D04}). To perform the integral for $\bm{k}_v$, we first consider three spheres 
\begin{eqnarray*}
 & & S_0: x^2 + y^2 +z^2 = k_{F}^{2}, \\
 & & S_1: (x+\vert \bm{q}' \vert \sin \theta  )^2 + y^2 + (z + \vert \bm{q}' \vert \cos \theta )^2 =k_{F}^{2}, \\
 & & S_2: x^2 + y^2 + (z + \vert \bm{q} + \bm{q}' \vert )^2 =k_{F}^{2}.
\end{eqnarray*}
Next, we define the circles $C_0$, $C_1$, and $C_2$ as the intersection of the plane $z=z_0$ with the sphere $S_0$, $S_1$, and $S_2$, respectively, 
\begin{eqnarray*}
 & & C_0: x^2 + y^2 = r_{0}^{2}, \\
 & & C_1: (x+\vert \bm{q}' \vert \sin \theta  )^2 + y^2 = r_{1}^{2}, \\
 & & C_2: x^2 + y^2 =r_{2}^{2}, 
\end{eqnarray*}
where
\begin{eqnarray*}
r_{0} &=& \sqrt{k_{F}^{2} -  z_{0}^{2} }, \\
r_{1} &=& \sqrt{k_{F}^{2} -  (z_{0} + \vert \bm{q}' \vert \cos \theta )^2}, \\
r_{2} &=& \sqrt{k_{F}^{2} - (z_{0} + \vert \bm{q} + \bm{q}' \vert )^2}, \\
z_{0} &=& \frac{m\omega}{\hbar \vert \bm{q} + \bm{q}' \vert} -\frac{\vert \bm{q} + \bm{q}' \vert}{2}.
\end{eqnarray*}
If we define $I_{ij}$ as the area of the intersection of the sphere $S_i$ with $S_j$, we obtain
\begin{eqnarray*}
 I_{00} &=& \pi \left( k_{F}^{2} -  z_{0}^{2} \right), \\
 I_{11} &=& \pi \left[ k_{F}^{2} -  (z_{0} + \vert \bm{q}' \vert \cos \theta)^2  \right], \\
 I_{22} &=& \pi \left[ k_{F}^{2} - (z_{0} + \vert \bm{q} + \bm{q}' \vert )^2) \right], \\
 I_{01} &=&  2(I_{01}^{-}+I_{01}^{+}), \\
 I_{12} &=&  2(I_{12}^{-}+I_{12}^{+}), 
\end{eqnarray*}
where
\begin{eqnarray*}
 I_{01}^{-} &=& \int_{x_{01}^{-}}^{x_{01}} \sqrt{(k_{F}^{2} -  z_{0}^{2}) - x^2} dx, \\
 I_{01}^{+} &=& \int_{x_{01}}^{x_{01}^{+}} 
 \sqrt{k_{F}^{2} -  (z_{0} + \vert \bm{q}' \vert \cos \theta)^2 - (x+\vert \bm{q}' \vert \sin \theta  )^2} dx, \\
 x_{01}^{-} &=& - \sqrt{k_{F}^{2} -  z_{0}^{2} }, \\
 x_{01}^{+} &=& - \vert \bm{q}' \vert \sin \theta 
 + \sqrt{k_{F}^{2} -  (z_{0} + \vert \bm{q}' \vert \cos \theta)^2}, \\
 x_{01} &=& - \frac{\vert \bm{q}' \vert + 2 z_0  \cos \theta}{2\sin \theta}, 
\end{eqnarray*}
and 
\begin{eqnarray*}
 I_{12}^{-} &=& \int_{x_{12}^{-}}^{x_{12}} 
 \sqrt{k_{F}^{2} - (z_{0} + \vert \bm{q} + \bm{q}' \vert )^2 - x^2} dx, \\
 I_{12}^{+} &=& \int_{x_{12}}^{x_{12}^{+}} 
 \sqrt{k_{F}^{2} -  (z_{0} + \vert \bm{q}' \vert \cos \theta)^2 - (x+\vert \bm{q}' \vert \sin \theta  )^2} dx, \\
 x_{12}^{-} &=& - \sqrt{k_{F}^{2} -  (z_{0} + \vert \bm{q} + \bm{q}' \vert)^{2} }, \\
 x_{12}^{+} &=& - \vert \bm{q}' \vert \sin \theta 
 + \sqrt{k_{F}^{2} -  (z_{0} + \vert \bm{q}' \vert \cos \theta)^2}, \\
 x_{12} &=& \frac{\vert \bm{q} + \bm{q}' \vert^2 + 2 z_0 \vert \bm{q} + \bm{q}' \vert 
 -2\vert \bm{q}' \vert z_0 \cos \theta - \vert \bm{q}' \vert^2}
 {2\vert \bm{q}' \vert \sin \theta}. 
\end{eqnarray*}
Here $x_{01}$ and $x_{12}$ are the solutions of simultaneous equations $C_0$ and $C_1$ and equations $C_1$ and $C_2$, respectively. The definite integrals of $I_{01}^{\pm}$ and $ I_{12}^{\pm}$ are calculated analytically. By using these expressions, the value of ${\rm Im} \tilde{D}_{2}^{\mathrm{T}(0-4)}(q,q',\sigma'';\omega)$ is expressed by
\begin{eqnarray}
{\rm Im} \tilde{D}_{2}^{\mathrm{T}(0-4)}(q,q',\sigma'';\omega)
 &=& \delta_{\sigma'\sigma} \delta_{\sigma''\sigma'} \delta_{\sigma\sigma''} 
 \left[- \frac{m\pi}{(2\pi)^3\hbar \vert \bm{q} + \bm{q}' \vert} \right] \nonumber\\
 &\times& \sum_{ij} \alpha_{ij} I_{ij},
\end{eqnarray}
where $\alpha_{ij}=-1, 0, 1$ whose value is determined by $\vert \bm{q} + \bm{q}'\vert$, $\vert \bm{q}' \vert$, and $\theta$.

\section{First order contribution}
In this section, we treat the electron-electron interaction Hamiltonian $H_1$ in Eq.~(\ref{eq:totalH}) as a perturbation and derive the first order contribution to $D_{2}^{\mathrm{T}} (x,x',x'';t-t')$ shown in Figs.~\ref{fig:diagram}($\rm{c}$) and \ref{fig:diagram}(d). The contribution from Fig.~\ref{fig:diagram}($\rm{c}$) is given by 
\begin{widetext}
\begin{eqnarray}
& & (-1) (-i) \left( -\frac{i}{\hbar} \right) \frac{1}{2} \int d x_1 \int d x_{1}' \int dt_1 \int dt_{1}'
 V(\bm{x}_1 - \bm{x}_{1}') \delta (t_1 - t_{1}')  \nonumber\\
&\times&  
  \left[ -i G^{(0)} (x_1, x'';t_1 - t') \right] 
  \left[ -i G^{(0)} (x, x_1;t - t_1) \right] 
  \left[ -i G^{(0)} (x', x;t- t^+) \right] \nonumber\\
  &\times&
  \left[ -i G^{(0)} (x_{1}', x';t_{1}'- t) \right] 
  \left[ -i G^{(0)} (x'', x_{1}';t'- t_{1}') \right], 
  \label{eq:F1}
\end{eqnarray}
\end{widetext}
where $(-1)$ denotes a closed loop and $(-i)$ comes from the definition of the time-ordered correlation function. The Fourier transformation in time of Eq.~(\ref{eq:F1}) yields
\begin{widetext}
\begin{eqnarray}
& & (-1) (-i)^6 \left( -\frac{i}{\hbar} \right) \frac{1}{2} \int d x_1 \int d x_{1}' 
 V(\bm{x}_1 - \bm{x}_{1}') 
 \delta_{\sigma_1\sigma''}  \delta_{\sigma\sigma_1} \delta_{\sigma'\sigma}
 \delta_{\sigma_{1}'\sigma'}  \delta_{\sigma''\sigma_{1}'}
  \left( i \sum_{\alpha}^{\rm occ} \phi_{\alpha\sigma'} (\bm{x}') \phi_{\alpha\sigma'}^* (\bm{x}) \right)  \nonumber\\
&\times&  \int \frac{d\omega_1}{2\pi} \int \frac{d\omega_2}{2\pi} 
   G^{(0)} (\bm{x}_1, \bm{x}''; \omega_1)  
   G^{(0)} (\bm{x}, \bm{x}_1; \omega_2) 
   G^{(0)} (\bm{x}_{1}', \bm{x}'; \omega_2 -\omega) 
   G^{(0)} (\bm{x}'', \bm{x}_{1}'; \omega_1 -\omega). 
   \label{eq:F2}
\end{eqnarray}
\end{widetext}
The Fourier transformation in space of Eq.~(\ref{eq:F2}) yields
\begin{widetext}
\begin{eqnarray}
& &  \delta_{\sigma''\sigma}  \delta_{\sigma'\sigma} \delta_{\sigma'\sigma''}
 \left( -\frac{1}{2\hbar} \right)
\int \frac{d \bm{k}_\alpha d \bm{k}_v}{(2\pi)^6} 
\theta (\vert \bm{k}_v + \bm{q} + \bm{q}' \vert -k_F) \theta (k_F - \vert \bm{k}_v \vert)
\theta (k_F - \vert \bm{k}_\alpha \vert) 
 \tilde{V}(\bm{k}_\alpha - \bm{k}_v - \bm{q}') \nonumber\\
 &\times& \left[ 
 \frac{\theta (k_F - \vert \bm{k}_\alpha - \bm{q}' \vert) \theta (\vert \bm{k}_\alpha + \bm{q} \vert - k_F) }
 {(\omega - \omega_{\bm{k}_v + \bm{q}+\bm{q}'} + \omega_{\bm{k}_v} +i\delta)
  (\omega - \omega_{\bm{k}_\alpha + \bm{q}} + \omega_{\bm{k}_\alpha - \bm{q}'} +i\delta)}
 -
 \frac{\theta (k_F - \vert \bm{k}_\alpha + \bm{q} \vert) \theta (\vert \bm{k}_\alpha - \bm{q}' \vert - k_F) }
 {(\omega - \omega_{\bm{k}_v + \bm{q}+\bm{q}'} + \omega_{\bm{k}_v} +i\delta)
  (\omega - \omega_{\bm{k}_\alpha + \bm{q}} + \omega_{\bm{k}_\alpha - \bm{q}'} -i\delta)}
  \right]  \nonumber\\
 &+& \delta_{\sigma''\sigma}  \delta_{\sigma'\sigma} \delta_{\sigma'\sigma''}\left( -\frac{1}{2\hbar} \right)
\int \frac{d \bm{k}_\alpha d \bm{k}_v}{(2\pi)^6} 
\theta (\vert \bm{k}_v - \bm{q} - \bm{q}' \vert -k_F) \theta (k_F - \vert \bm{k}_v \vert)
\theta (k_F - \vert \bm{k}_\alpha \vert) 
 \tilde{V}(\bm{k}_\alpha - \bm{k}_v + \bm{q}) \nonumber\\
 &\times& \left[ 
 - \frac{\theta (k_F - \vert \bm{k}_\alpha - \bm{q}' \vert) \theta (\vert \bm{k}_\alpha + \bm{q} \vert - k_F) }
 {(\omega - \omega_{\bm{k}_v} + \omega_{\bm{k}_v - \bm{q}-\bm{q}'} -i\delta)
  (\omega - \omega_{\bm{k}_\alpha + \bm{q}} + \omega_{\bm{k}_\alpha - \bm{q}'} +i\delta)}
 +
 \frac{\theta (k_F - \vert \bm{k}_\alpha + \bm{q} \vert) \theta (\vert \bm{k}_\alpha - \bm{q}' \vert - k_F) }
 {(\omega - \omega_{\bm{k}_v} + \omega_{\bm{k}_v-\bm{q}-\bm{q}'} -i\delta)
  (\omega - \omega_{\bm{k}_\alpha + \bm{q}} + \omega_{\bm{k}_\alpha - \bm{q}'} -i\delta)}
  \right]. \nonumber\\
  \label{eq:F3}
\end{eqnarray}
\end{widetext}
Since we consider $\omega>0$, the imaginary part of Eq.~(\ref{eq:F3}) is 
\begin{eqnarray}
 \delta_{\sigma''\sigma}  \delta_{\sigma'\sigma} \delta_{\sigma'\sigma''}
 (I_1 + I_2 + I_3 + I_4),
\end{eqnarray}
where
\begin{widetext}
\begin{eqnarray}
I_1 &=& \left( -\frac{1}{2\hbar} \right)
\int \frac{d \bm{k}_\alpha d \bm{k}_v}{(2\pi)^6} 
\theta (\vert \bm{k}_v + \bm{q} + \bm{q}' \vert -k_F) \theta (k_F - \vert \bm{k}_v \vert)
\theta (k_F - \vert \bm{k}_\alpha \vert) 
 \tilde{V}(\bm{k}_\alpha - \bm{k}_v - \bm{q}') \nonumber\\
 &\times& (-\pi)
  {\cal P} \left( 
 \frac{\theta (k_F - \vert \bm{k}_\alpha - \bm{q}' \vert) \theta (\vert \bm{k}_\alpha + \bm{q} \vert - k_F) }
 {\omega - \omega_{\bm{k}_v + \bm{q} + \bm{q}' } + \omega_{\bm{k}_v}}
 \right)
 \delta (  \omega - \omega_{\bm{k}_\alpha + \bm{q}} + \omega_{\bm{k}_\alpha - \bm{q}'}), 
\end{eqnarray}
\begin{eqnarray}
I_2 &=& \left( -\frac{1}{2\hbar} \right)
\int \frac{d \bm{k}_\alpha d \bm{k}_v}{(2\pi)^6} 
\theta (\vert \bm{k}_v + \bm{q} + \bm{q}' \vert -k_F) \theta (k_F - \vert \bm{k}_v \vert)
\theta (k_F - \vert \bm{k}_\alpha \vert) 
 \tilde{V}(\bm{k}_\alpha - \bm{k}_v - \bm{q}') \nonumber\\
 &\times& (-\pi)
{\cal P} \left( 
 \frac{\theta (k_F - \vert \bm{k}_\alpha - \bm{q}' \vert) \theta (\vert \bm{k}_\alpha + \bm{q} \vert - k_F) }
 { \omega - \omega_{\bm{k}_\alpha + \bm{q}} + \omega_{\bm{k}_\alpha - \bm{q}'}}
 \right)
 \delta ( \omega - \omega_{\bm{k}_v + \bm{q}+\bm{q}'} + \omega_{\bm{k}_v}), 
\end{eqnarray}
\begin{eqnarray}
I_3 &=& \left( -\frac{1}{2\hbar} \right)
\int \frac{d \bm{k}_\alpha d \bm{k}_v}{(2\pi)^6} 
\theta (\vert \bm{k}_v + \bm{q} + \bm{q}' \vert -k_F) \theta (k_F - \vert \bm{k}_v \vert)
\theta (k_F - \vert \bm{k}_\alpha \vert) 
 \tilde{V}(\bm{k}_\alpha - \bm{k}_v - \bm{q}') \nonumber\\
 &\times& (+\pi)
  {\cal P} \left( 
 \frac{ \theta (k_F - \vert \bm{k}_\alpha + \bm{q} \vert) \theta (\vert \bm{k}_\alpha - \bm{q}' \vert - k_F)}
 { \omega - \omega_{\bm{k}_\alpha + \bm{q}} + \omega_{\bm{k}_\alpha - \bm{q}'}}
 \right)
 \delta ( \omega - \omega_{\bm{k}_v +\bm{q}+\bm{q}'} + \omega_{\bm{k}_v}), 
\end{eqnarray}
\begin{eqnarray}
I_4 &=& \left( -\frac{1}{2\hbar} \right)
\int \frac{d \bm{k}_\alpha d \bm{k}_v}{(2\pi)^6} 
\theta (\vert \bm{k}_v - \bm{q} - \bm{q}' \vert -k_F) \theta (k_F - \vert \bm{k}_v \vert)
\theta (k_F - \vert \bm{k}_\alpha \vert) 
 \tilde{V}(\bm{k}_\alpha - \bm{k}_v + \bm{q}) \nonumber\\
 &\times& (+\pi)
 {\cal P} \left( 
 \frac{ \theta ( \vert \bm{k}_\alpha + \bm{q} \vert- k_F) \theta (k_F - \vert \bm{k}_\alpha - \bm{q}' \vert )}
 { \omega - \omega_{\bm{k}_v } + \omega_{\bm{k}_v-\bm{q}-\bm{q}'} }
 \right)
 \delta ( \omega - \omega_{\bm{k}_\alpha + \bm{q}} + \omega_{\bm{k}_\alpha - \bm{q}'}). 
\end{eqnarray}
\end{widetext}
The integral for $\bm{k}_\alpha$ and $\bm{k}_v$ is computed by using the standard Monte Carlo approach ($10^8$ sampling points were used.). To obtain the contribution from Fig.~\ref{fig:diagram}(d), the replacement $\bm{q}\leftrightarrow \bm{q}'$ is needed in these expressions of $I_j \ (j=1,2,3,4)$. Finally, we can calculate ${\rm Im} \tilde{D}_{2}^{\mathrm{R}}(q,q',\sigma'';\omega)$ appeared in Eqs.~(\ref{eq:I_0}) and (\ref{eq:I_1}).



\begin{thebibliography}{99}
\bibitem{note1} It was confirmed that the use of the Hubbard-type formula described in Ref.~\cite{ichimaru} yields the same qualitative results.
\bibitem{note3} The relative momentum is important to study the exciton dispersion. The dispersion relation of four molecular crystals has been investigated recently.\cite{cudazzo1,cudazzo2}
\end{thebibliography}
\end{document}